\documentclass[prl,aps,showpacs,twocolumn,floatfix]{revtex4}
\usepackage{graphicx}

 \def\ba#1{\begin{array}{#1}}
 \def\ea{\end{array}}
 \def\be{\begin{equation}}
 \def\ee{\end{equation}}
\def\bq{\begin{equation}}
\def\eq{\end{equation}}
 \def\br{\begin{eqnarray}}
 \def\er{\end{eqnarray}}

\begin{document}

\title{   Non-degenerate shell-model effective interactions from\\ 
 the Okamoto-Suzuki and Krenciglowa-Kuo iteration methods }
\author{Huan\ Dong$^{1}$, T.\ T.\ S.\ Kuo$^{1}$ and J.\ W.\ Holt$^{2}$}
\affiliation{$^1$Department of Physics, State University New York at Stony 
Brook\\ Stony Brook, New York 11794}
\affiliation{$^2$Physik Department, Technische Universit\"at M\"unchen,
D-85747 Garching, Germany}
\date{\today}

\begin{abstract}
We present calculations of shell-model effective interactions  for both
degenerate and non-degenerate model spaces  using
the  Krenciglowa-Kuo (KK) and the extended Krenciglowa-Kuo iteration method
recently developed by  Okamoto, Suzuki {\it et al.} (EKKO). 
The starting point is the low-momentum nucleon-nucleon interaction 
$V_{low-k}$ obtained from the N$^3$LO 
chiral two-nucleon interaction. The model spaces spanned by the
$sd$ and $sdpf$ shells are both considered.
With a  solvable model, we show 
that both the KK and EKKO methods are convenient
for deriving the effective interactions for non-degenerate 
model spaces.  The EKKO method is especially 
desirable in this situation since the vertex function 
$\hat Z$-box employed therein is well behaved 
while the corresponding vertex function $\hat Q$-box employed in the 
Lee-Suzuki (LS) and KK 
methods may have singularities. The converged shell-model 
effective interactions
given by the EKKO and KK methods are equivalent, although the former
method is considerably more efficient. 
The  degenerate $sd$-shell effective interactions 
given by the LS method are practically identical to those from 
the EKKO  and KK methods. Results of the $sd$ one-shell and $sdpf$ two-shell
 calculations for
$^{18}$O, $^{18}$F, $^{19}$O and $^{19}$F using the EKKO effective
interactions are compared, and the importance of 
 the shell-model three-nucleon forces is discussed. 
\end{abstract}

\maketitle


\section{I. Introduction}

 The nuclear shell model  has provided a very successful framework for 
describing the properties of a wide range of nuclei.
This framework is basically an effective theory 
\cite{ko90,jensen95,coraggio09}, 
corresponding to reducing the
full-space nuclear many-body problem  to a model-space
one with effective Hamiltonian  $PH_{eff}P$=$PH_0P+PV_{eff}P$, 
where $H_0$ is the single-particle (s.p.) Hamiltonian 
and $P$ represents the projection
operator for the model space which is usually chosen 
to be a small shell-model space
such as the $sd$ shell outside of an $^{16}{\rm O}$ closed core.   
The effective interaction $V_{eff}$  plays a central role
in this nuclear shell model approach, and its choice and/or determination
have been extensively studied, see e.g.\ 
\cite{brownwild88,brownrich06,jensen95,coraggio09}. As discussed in these 
references, $V_{eff}$ may be determined using either an empirical approach
where it is required to reproduce  selected experimental data or 
a microscopic one where $V_{eff}$
is derived  from realistic nucleon-nucleon
(NN) interactions using many-body methods. The folded-diagram theory
\cite{ko90,jensen95,coraggio09} is a commonly used such method for 
the latter. Briefly speaking, in this theory
$V_{eff}$ is given
 as a folded-diagram series \cite{ko90,jensen95,coraggio09,klr}
\bq
V_{eff}= \hat{Q} - \hat{Q}^{'}\int\hat{Q}
+ \hat{Q}^{'}\int\hat{Q}\int\hat{Q} - \hat{Q}^{'}\int\hat{Q}\int\hat{Q}
\int\hat{Q}\cdots,
\label{veff}
\eq
where $\hat Q$ represents a so-called $\hat Q$-box, which may be
 written as
\bq
\hat Q(\omega)=[PVP+PVQ\frac{1}{\omega -QHQ}QVP]_{L}.
\label{qboxe}
\eq
Here $V$ represents the NN interaction and $\omega$ is the so-called
starting energy which will be explained later (section II). 
Thus  from the NN interaction $V$
we can in principle calculate the $\hat Q$-box and thereby 
the effective interaction $V_{eff}$. Note that we use $Q$, 
without hat, to denote the $Q$-space projection operator.
($P+Q=1$.) Note also that the $\hat Q$-box is an irreducible vertex function 
where the intermediate states between any two vertices must 
belong to the $Q$ space. As indicated by the subscript $L$ in eq.\ (\ref{qboxe}), 
the $\hat Q$-box contains valence linked 
diagrams only, such as the 1st- and 2nd-order
$\hat Q$-box diagrams for $^{18}{\rm O}$ and $^{18}{\rm F}$ shown in Fig.\ 1.
 The $\hat Q'$-box of
eq.\ (\ref{veff}) is defined as ($\hat Q - PVP$), namely $\hat Q'$ begins with 
diagrams 2nd-order in $V$. The above folded-diagram formalism has been
employed in microscopic derivations of shell model effective interactions
for a wide range of nuclei \cite{jensen95,coraggio09}.

 In the present work we would like to explore an extension of the well-known
methods for computing the folded diagram series in eq.\ (\ref{veff}). 
The Lee-Suzuki (LS) \cite{lesu,sule,suzu94} iteration scheme has been commonly 
used in previous microscopic calculations of shell model effective interactions 
\cite{jensen95,coraggio09}. 
Here we would like to employ two different methods,
the Krenciglowa-Kuo (KK) iteration method \cite{kren,kuo95}
 and the newly developed extended
 Krenciglowa-Kuo iteration method of 
Okamoto, Suzuki, Kumagai and Fujii (EKKO)\cite{okamoto10},  mainly for the
purpose of calculating the shell-model efective interactions
for non-degenerate model spaces. 
 As we shall discuss later, it is not convenient
to use the LS method  for calculating the effective interactions for
 non-degenerate  model spaces such as the 
 $sdpf$ two-shell case, while both the EKKO and the KK
 methods can be conveniently applied in this situation. 
The EKKO method has an additional advantage. When the $P$- and $Q$-space are not 
adequately separated from each other,
the $\hat Q$-box employed in the  KK method may have singularities,
causing difficulty for its iterative solution.
An essential and interesting difference between the EKKO and 
  KK methods is that
 the EKKO method employs the vertex function $\hat Z$-box (to be
defined in section II) 
while in the latter the vertex function $\hat Q$-box is used.
 This simple replacement (of $\hat Q$ by $\hat Z$) 
has an  important advantage in circumventing the singularities
mentioned above. As we shall discuss later, both the EKKO and  
KK methods may provide a  suitable 
 framework for calculating shell-model effective interactions for large
non-degenerate model spaces which may be needed for describing
exotic nuclei with large neutron excess. 

The organization of the present paper is as follows. 
In section II we shall
describe a non-degenerate version of the EKKO method \cite{okamoto10} and 
 how we apply it and the KK method \cite{kren,kuo95} 
to shell-model effective interactions. A comparison of these two
methods with the LS scheme \cite{lesu,sule}  will be made. 
 Our results will be presented and discussed in
 section III.  We shall first perform a sequence of model 
calculations comparing the EKKO, KK and LS iteration methods
for both degenerate and non-degenerate model spaces. 
The use of the EKKO and KK methods in calculating the
effective interactions for non-degenerate model spaces
will be emphasized.
Starting from the $V_{low-k}$ interaction 
\cite{bogner01,bogner02,bogner03,coraggio09}
derived from the chiral
N$^3$LO potential \cite{idaho}, the LS, KK and 
EKKO methods will all be used to calculate
the degenerate $sd$ one-shell effective interactions, a main purpose being
 to check if the results given by
the commonly used LS method agree with the KK and EKKO ones.
The KK and EKKO methods will then be employed to calcualte the non-degenerate
 $sdpf$ two-shell effective interactions. The matrix elements of the
above degenerate and non-degenerate interactions will be compared. 
The low-energy
spectra of $^{18}$O, $^{18}$F, $^{19}$O and $^{19}$F given by these
interactions will be discussed.
A summary and conclusion will be presented in section IV.

\begin{figure}
\scalebox{0.3}{
\includegraphics[angle=-90]{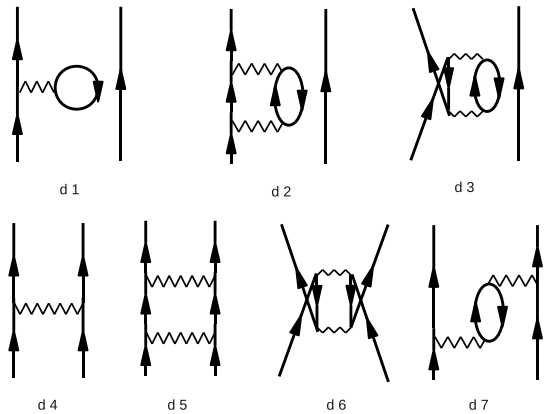}
}
\caption{Low-order diagrams constituting the $\hat Q$-box.}
\end{figure}


\section{II. Formalism}
\label{formalism}

In this section, we shall describe and discuss the KK
\cite{kren,kuo95} and  the EKKO
\cite{okamoto10} iteration methods and their application to 
microscopic  calculations of shell-model
effective interactions. These methods, to our knowledge, have not yet
been employed in such calculations. Let us begin with a brief 
review of the LS \cite{lesu,sule} and KK iteration methods. 
Consider first the degenerate LS method
where the model space is  degenerate, namely  $PH_0P=W_0$, $W_0$ being
a constant. In terms of the
$\hat Q$-box of Eq.(2), the effective interactions $R_n$ are calculated
iteratively by \cite{lesu,sule}
\bq
R_{1} = \hat Q(W_0),
\eq
\bq
R_{2}= \frac{1}{1- \hat Q_1}
 \hat{Q}(W_0), 
\eq
and for $n>2$
\bq
R_{n} = \frac{1}{1-\hat{Q}_1 -\sum_{m=2}^{n-1} \hat{Q}_m \prod_{k=n-m+1}
^{n-1} R_{k} } \hat{Q}(W_0),
\eq
where $\hat Q_m$ is proportional to the $m{\rm th}$ derivative of $\hat Q$:
\bq
\hat{Q}_m = \left . \frac{1}{m!}\frac{d^m \hat{Q}}{d\omega^m}\right |_{\omega=W_0}.
\eq
The effective interaction is given by the converged $R_n$, namely
\bq
V_{eff}=R_{n+1}=R_n.
\eq

There is a practical difficulty for the above iteration method.
In actual shell-model calculations, it is usually not possible to calculate
the vertex function $\hat Q$-box exactly;  thus it is a common practice
to evaluate it with some low-order approximation and calculate
the derivatives $\hat Q_m$ numerically.  At higher orders this becomes 
increasingly difficult, and therefore such calculations are usually
limited to low orders in the iteration. However, as we shall demonstrate in 
section IIIa, low-order LS iterations are often
not accurate when the $P$- and $Q$-spaces are strongly coupled. 

 The above  degenerate LS iteration method can be generalized to 
 a non-degenrate one \cite{suzu94}, namely $PH_0P$ being
non-degenerate.  In this situation we need not only the 
$\hat Q$-box of eq.\ (\ref{qboxe}) but also a generalized 
$\hat{Q}$-box defined by 
\begin{equation}
 \hat{Q}_n(\epsilon_1 \epsilon_2 \cdots \epsilon_{n+1})
 = \\ (-1)^n [PVQ g^Q_1g^Q_2\cdots  g^Q_{n+1}QVP]_L.
\end{equation}
with
\begin{equation}
g^Q _{i}\equiv 
 \frac{1}{\epsilon_i-QHQ}.
\end{equation}
This generalized $\hat Q$-box is defined for $n \ge 1$, and 
$\epsilon_i$ is defined by 
$PH_0P\phi_i=\epsilon_i \phi_i$ where
$P=\sum_{m\leq d} |\phi_m \rangle \langle \phi _m |$.
The dimension of the $P$-space is labelled $d$.
Note that only valence-linked diagrams are retained in 
$\hat Q _n(\epsilon_1 \cdots \epsilon_{n+1})$, 
as indicated by the subscript L.

With the above definitions, the effective interaction given by the
 non-degenerate  LS iteration method is given as \cite{suzu94}
 \begin{eqnarray}
 R_1&=&\sum_{\alpha}\hat{Q}(\epsilon_{\alpha})P_{\alpha},
 \nonumber\\
 R_2&=&\sum_{\alpha}
 \left[1-\sum_{\beta}
 \hat{Q}_1\left(\epsilon_{\alpha \beta}\right)
 P_{\beta}\right]^{-1}
 \hat{Q}(\epsilon_{\alpha})P_{\alpha},
 \nonumber\\
 R_3&=&\sum_{\alpha}
 \left[1-\sum_{\beta}\hat{Q}_1\left(\epsilon_{\alpha \beta} \right)
 P_{\beta} 
   -\sum_{\beta \gamma}\hat{Q}_2\left(\epsilon_{\alpha
\beta \gamma} \right)P_{\beta}R_2P_{\gamma}
 \right]^{-1}  \nonumber \\
&& \times \hat{Q}(\epsilon_{\alpha})P_{\alpha},
 \nonumber \\
 &\cdots&   
\end{eqnarray}
with $\epsilon_{\alpha \beta}\equiv (\epsilon_{\alpha }+\epsilon_{\beta})$,
 $\epsilon_{\alpha \beta \gamma}\equiv (\epsilon_{\alpha }+\epsilon_{\beta}
+\epsilon _{\gamma}),~\cdots$.
In the above equations $P_m = |\phi_m \rangle \langle \phi _m |$. 
When convergent, we have $V_{eff}=R_{n+1}=R_n$. The above non-degenerate
LS method can be used to calculate the effective interactions for e.g.,
the non-degenerate $0p$-shell \cite{coraggio05} 
or the two-shell $sdpf$ model space. But this method is rather
complicated for computations, and this has hindered its application
to microscopic calculations of shell-model effective interactions.

 We now describe some details of the non-degenerate KK and EKKO methods.
 The KK iteration method  was originally developed for
 model spaces which are degenerate \cite{kren}. A non-degenerate KK iteration
method was later formulated  \cite{kuo95}, with the effective interaction 
 given by the following iteration methods.
 Let the effective interaction for the {\it i}th iteration be
$V_{eff}^{(i)}$ and the corresponding eigenvalues $E$ and eigenfunctions
$\chi$ be given by
\begin{equation}
[PH_0P +V_{eff}^{(i)}]\chi_m^{(i)}=E_m^{(i)} \chi_m^{(i)}.
\end{equation}
Here  $\chi _m$ is the $P$-space projection of the full-space eigenfunction
$\Psi _m$, namely $\chi _m= P\Psi _m$.
The effective interaction for the next iteration is then
\begin{equation}
V_{eff}^{(i+1)}=\sum_m[PH_0P+\hat Q(E_m^{(i)})]
| \chi_m^{(i)}\rangle \langle \tilde{\chi} _m^{(i)}| -PH_0P,
\end{equation}
where the bi-orthogonal states are defined by
\begin{equation}
 \langle \tilde \chi _m |\chi _{m'} \rangle = \delta _{m,m'}. 
\end{equation}
Note that in the above $PH_0P$ is non-degenerate. The converged eigenvalue
$E_m$ and eigenfunction $\chi _m$ satisfy the $P$-space 
self-consistent condition
\bq
(E_m(\omega)-H_0)\chi_m=\hat Q(\omega)\chi _m,~ \omega=E_m(\omega).
\eq 

To start the iteration, we use
\begin{equation}
V_{eff}^{(1)}=\hat Q(\omega _0)
\end{equation}
where $\omega _0$ is a starting energy chosen to be close to $PH_0P$.
 The converged KK effective interaction is given by 
$V_{eff}=V_{eff}^{(n+1)}=V_{eff}^{(n)}$. 
When convergent, the  resultant $V_{eff}$  is independent
of $\omega _0$, as  it is the states
with maximum $P$-space overlaps which are selected by the KK method \cite{kren}. 
 We shall discuss  this feature
later in section III using a solvable model. 
The above non-degenerate
KK method is numerically more convenient than the non-degenerate LS
method.

The diagrams of Fig.\ 1 have both one-body (d1,d2,d3) and two-body (d4,d5,d6,d7)
diagrams. When we calculate nuclei with two valence nucleons such as
$^{18}{\rm O}$ and $^{18}{\rm F}$, all seven diagrams are included in the 
$\hat Q$-box. But for nuclei with one valence nucleon such as $^{17}{\rm O}$, 
we deal with the 1-body $\hat S$-box which is approximated by 
the sum of diagrams d1, d2 and d3. The 1-body effective interaction
is given by a similar KK iteration
\begin{equation}
S_{eff}^{(i+1)}=\sum_m[PH_0P+\hat S(E_m^{(i)})]
| \chi_m^{(i)}\rangle \langle \tilde{\chi} _m^{(i)}| -PH_0P.
\end{equation}
 Denoting its converged value as $S_{eff}$, the model-space
 s.p.\ energy $\epsilon _m^{eff}$ is given by $P_m(H_0+S_{eff})P_m$.
 By adding and then subtracting $S_{eff}$, we can rewrite Eq.(14) as
\bq
(E_m(\omega)-H_0^{eff})\chi_m=[\hat Q(\omega)-S_{eff}]\chi _m,
~ \omega=E_m(\omega),
\eq
with $H_0^{eff}=H_0 + S_{eff}$.  
In most shell model calculations \cite{brownwild88,brownrich06}, 
one often uses the experimental s.p.\ energies. 
This treatment for the s.p.\ energies is in line with the above subtraction 
procedure, as  $P(H_0+S_{eff})P$ represents  the physical s.p.\ energy
which in principle can be extracted from experiments. 
In the present work we shall use the experimental s.p.\ energies
for the model-space orbits together with the $V_{eff}$ derived 
from ($\hat Q -S_{eff}$). 
A similar subtraction procedure has also been
employed in the LS calculations \cite{jensen95,coraggio09} where
the all-order sum of the one-body diagrams was subtracted from the
calculation of the effective interaction. 

In several aspects, the above KK method provides a more desirable
framework for effective interaction calculations than the commonly
used LS method.  The KK method is more convenient for
non-degenerate model spaces than the LS method, and the KK method
does not require the calculation of high-order derivatives of the 
$\hat Q$-box, which may be necessary in a converged LS calculation.
The KK method has, however, a shortcoming when applied to calculations
with extended model space such as the two-shell $sdpf$ one. For example,
certain 2nd-order diagrams for this case may diverge, resulting 
in an ill-defined $\hat Q$-box.  It is remarkable
that these potential divergences can be circumvented by the recently
proposed EKKO method of Okamoto et al.\ \cite{okamoto10}.
In this method, the vertex function $\hat Z$-box is employed. It is
related to the $\hat Q$-box by
\begin{equation}
\hat Z(\omega)=\frac{1}{1-\hat Q_1(\omega)}[\hat Q(\omega)
-\hat Q_1 (\omega)P(\omega-H_0)P],
\end{equation}
where $\hat Q_1$ is the first-order derivative of the $\hat Q$-box.
The $\hat Z$-box considered by Okamoto {\it et al.} \cite{okamoto10} is for
degenerate model spaces ($PH_0P=W_0$), while we consider here a more general
case with non-degenerate  $PH_0P$.
An important property of the above $\hat Z$-box is that it is finite when
the $\hat Q$-box is singular (has poles). Note that $Z(\omega)$ satisfies
\bq
\hat Z(\omega) \chi _m =\hat Q(\omega)\chi_m ~{\rm at}~ \omega=E_m(\omega).
\eq 
The iteration method for determining the effective interaction from the
$\hat Z$-box is quite similar to that for the $\hat Q$-box. Suppose
the effective interaction for the $i$th iteration is $V_{eff-Z}^{(i)}$.
The corresponding eigenfunction $\chi$ and eigenvalues $ E^Z$ are determined
by
\begin{equation}
[PH_0P +V_{eff-Z}^{(i)}]\chi_m^{(i)}= E_m^{Z(i)} \chi_m^{(i)}.
\end{equation}
The effective interaction for the next iteration is
\begin{equation}
V_{eff-Z}^{(i+1)}=\sum_m[PH_0P+\hat Z( E_m^{Z(i)})]
| \chi_m^{(i)}\rangle \langle \tilde{\chi} _m^{(i)}| -PH_0P,
\end{equation}
Although $\hat Q(\omega)$ and $\hat Z(\omega)$ are generally different, 
 it is interesting that the converged eigenvalues $E_m$ 
of $P(H_0+V_{eff})P$ 
and the corresponding ones $ E_m^Z $ of $P(H_0+V_{eff-Z})P$  are 
 both exact eigenvalues of the full-space Hamiltonian $H=H_0+V$,
which can be seen from eqs.\ (14), (18) and (19). 
Note, however,  the KK and EKKO methods may reproduce different eigenvalues
of the full-space Hamiltonian $H$.  This aspect together with 
some other comparisons
of these methods will be discussed in section IIIa, using a simple
solvable model.

For the degenerate case, Okamoto {\it et al}.\ \cite{okamoto10}
have shown that
$\frac{d E_m^Z(\omega)}{d\omega}$ =0 at the self-consistent
point $\omega= E_m^Z(\omega)$. 
As outlined below, we have found that this result also holds
for the case of non-degenerate $PH_0P$.
From eq.\ (18), we have
\begin{eqnarray}
 \frac{dZ}{d\omega}&=&
\frac{2}{1-\hat Q_1} \hat Q_2 \frac{1}{1-\hat Q_1}
[\hat Q-\hat Q_1(\omega-H_0)] \\  \nonumber
& & -\frac{2}{1-\hat Q_1}\hat Q_2 (\omega -H_0). 
\end{eqnarray}
Then from eqs.\ (14) and (19) we have
\begin{equation} 
[\frac{dZ(\omega)}{d\omega}]_{\omega= E_m^Z}|\chi _m \rangle=0,
\end{equation}
and
\begin{equation}
[\frac{d  E_m^Z(\omega)}{d \omega}]_{\omega= E_m^Z}=0.
\end{equation}
This is a useful result; it states that at any self-consistent point
the eigenvalues $ E_m^Z(\omega)$ varies `flatly' with $\omega$, a feature
certainly helpful to iterative calculations. In section III, we shall check
this feature numerically.

\section{III. Results and discussion}
\subsection{IIIa. Model calculations comparing the LS, KK and EKKO
methods}

In this section we shall study the above iteration methods
by way of a simple matrix model,
similar to the one employed in \cite{okamoto10}. We consider 
a 4-dimensional matrix Hamiltonian  $H=H_0+H_1$ where
\begin{equation}
H_0 =
\left[ \begin{array}{cc}
PH_0P & 0  \\
0 & QH_0 Q  
\end{array} \right ]
\end{equation}
and
\begin{equation}
PH_0P =
\left[ \begin{array}{cc}
\varepsilon _{p1} & 0  \\
0 & \varepsilon _{p2}  
\end{array} \right ];~
QH_0Q =
\left[ \begin{array}{cc}
\varepsilon _{q1} & 0  \\
0 & \varepsilon _{q2}  
\end{array} \right ].
\end{equation}
The interaction Hamiltonian has a strength parameter $x$, namely
\begin{equation}
H_1 =
\left[ \begin{array}{cc}
PH_1P & PH_1Q  \\
QH_1 P & QH_1Q 
\end{array} \right ],
\end{equation}
with
\begin{eqnarray}
PH_1P &=&
\left[ \begin{array}{cc}
0 & 5x  \\
5x & 10x  
\end{array} \right ]; \nonumber \\
PH_1Q &=& QH_1P=
\left[ \begin{array}{cc}
0 & 8x  \\
8x & 0 
\end{array} \right ]; \nonumber \\
QH_1Q &=&
\left[ \begin{array}{cc}
-5x & x  \\
x &-5x 
\end{array} \right ].
\end{eqnarray}

As discussed in section II, both the KK and EKKO iteration methods are rather
convenient for non-degenerate model spaces. We would like to check 
 this feature by carrying out some calculations using the above model.
We consider two unperturbed Hamiltonians, given by
 $(\varepsilon _{p1},~
 \varepsilon _{p2},~
 \varepsilon _{q1},~
 \varepsilon _{q2})$=(0,6,4,9) and (0,0,4,9). The $PH_0P$  
parts of them are, respectively, non-degenerate and degenerate.

In the first three entries of Table I,  some results for the
 $PH_0P$=(0,6) case are presented. Here $E_n$ are the exact
eigenvalues of the full Hamiltonian, with their model-space
overlaps denoted by  $(n|P|n)$. The entries $E_{KK}$ and $E_{EKKO}$ are the 
eigenvalues generated respectively by the KK and EKKO iteration methods.
Not only is the above $PH_0P$ non-degenerate but its spectrum
intersects that of $QH_0Q$. One would expect that this $PH_0P$ may cause difficulty
for the above iteration  methods. But as indicated in Table I, both the 
non-degenerate KK and the non-degenerate EKKO iteration
methods work remarkably well. 
Note that the interaction used here  is rather strong (x=0.6), 
and both methods still work well, converging to values of $E_n$ which are 
quite far from $PH_0P$.

Some results for the degenerate case of $PH_0P$=(0,0)  are listed in 
the last two entries of Table I.
Here we have performed calculations using the degenerate LS method
 through 5th order iteration (i.e.\ in Eq.(7)
we  use $V_{eff}=R_5$). As shown, the results so obtained are not
in good agreement with the exact results. This suggests that
 low-order LS iteration method may often be inadequate, and one needs
higher-order iterations to obtain accurate results.

It is known that the KK iteration method converges
to the states with maximum $P$-space overlaps \cite{kren}, while the
LS method converges to the states of lowest energies \cite{lesu,sule}.
We have found that for many cases the EKKO method
also converges to states of maximum $P$-space overlaps.
 As listed in the third entry 
of Table I, both the KK and EKKO methods converge to states of energies
$E_n$= -3.51 and 14.53 whose $P$-space probabilities are relatively
0.70 and 0.86.  
We have also found that the EKKO and KK iteration methods can converge
to different states. An example is the result shown in the last part
of Table I, where the EKKO method converges to states of energy
($P$-space probability) -1.45 (0.87) and 0.91 (0.46), while the
states of maximum $P$-space overlap are those with energies -1.45 and 
5.25. Note that for this case the EKKO method is clearly
more accurate than the KK method.

We have noticed that for a number of cases  the EKKO method converges well
but not so for the KK method. This is largely because these two methods
treat the singularities of the $\hat Q$-box differently. To see this,
let us perform a graphical solution for the x=0.60 case of Table I.
Using the parameters of this case, we calculate and plot in
Fig.\ 3 both $E_m^Q(\omega)$ and $E_m^Z(\omega)$ which are respectively the
eigenvalues of
$P[H_0+\hat Q(\omega)]P$ and $P[H_0+\hat Z(\omega)]P$. As discussed in section
II, they have identical self-consistent solutions, namely 
$\omega=E_m^Q(\omega)=E_m^Z(\omega)\equiv E_m$ where $E_m$ is the eigenvalue
of the full-space Hamiltonian.
Recall that $\hat Q$ and $\hat Z$ are given respectively by eqs.\ (2) and (18).
As shown in the figure, the curves of $E^Q$ and $E^Z$ do have identical
self-consistent solutions as marked by the common intersection points
$E_1$, $E_2$, $E_3$ and $E_4$. Note that
the above two curves are distinctly
different from each other, particularly in the vicinity
of the poles  (marked by the vertical lines through $F_1$ and $F_2$) 
 of the $\hat Q$-box. There $E^Q(\omega)$ is discontinuous,
diverging oppositely before and after the pole, while $E^Z(\omega)$ remains
 continuous throughout. This clearly helps the convergence of the
$\hat Z$-box iteration method: The $\hat Z$-box iteration proceeds
along a continuous $E^Z(\omega)$ curve, while the $\hat Q$-box iteration
often does not converge as it may bounce back and forth 
 across the discontinuity.

 As seen from eq.\ (18),  the $\hat Z$-box method has `false' solutions
at $E_q^Z(\omega)=\omega = \mu_q \equiv F_q$, where $\mu_q$ are poles
of the $\hat Q$-box.  
These solutions are marked 
in  Fig.\ 3 as $F_1$ and $F_2$. These false solutions can be readily recognized
and discarded. As given in Eq.(24), we have at self-consistent points 
 $\frac{dE_m^Z}{d\omega}=0$. As shown in Fig. 3, the slopes
of $E^Z$ do satisfy the above condition at the self-consistent
points $E_1$ to $E_4$, but not so at the false points $F_1$ and $F_2$.


\begin{table}
\caption{Results  of model calculations using 
the LS, KK and EKKO iteration methods. See text for other explanations.}
\begin{center}
\begin{tabular}{ccccc}\hline \hline
 $H_0$= (0,6,4,9)&x=0.10&&&\\
 $E_n$ &   -0.110705 & 3.328164 & 7.203243 & 8.579299\\
 $(n|P|n)$ &     0.991108 & 0.044558&  0.953368 & 0.010966\\
 $E_{KK}$  &  -0.110705&  7.203243 &&\\
 $E_{EKKO}$&  -0.110705&  7.203242&& \\ \hline
 &  x =0.30 &&& \\
 $E_n$ &   -0.974184 & 1.808716 & 8.080888& 10.084581\\
 $(n|P|n)$ &     0.906155&  0.111257&  0.113523 & 0.869066\\
 $E_{KK}$  &  -0.974185& 10.084580 &&\\
 $E_{EKKO}$ & -0.974185 &10.084581 &&\\ \hline
  &  x =0.60 &&& \\
 $E_n$  &  -3.510377& -0.286183&  8.265089& 14.531471 \\
 $(n|P|n)$  &    0.709531&  0.189318&  0.232422&  0.868729\\
 $E_{KK}$ &   -3.510377 &14.531472&&\\
 $E_{EKKO}$ & -3.510363 & 14.531472 && \\ \hline \hline
  $H_0$=(0,0,4,9)& x= 0.10 &&&\\
 $E_n$   & -0.296201 & 0.982861 & 3.736229 & 8.577111 \\
$ (n|P|n)$     & 0.985416 & 0.922993 & 0.082811 & 0.008780 \\
 $E_{LS}$ & -0.314591 & 0.872176&&\\
 $E_{KK}$  &  -0.296201 & 0.982860&&\\
 $E_{EKKO}$&  -0.296201 & 0.982860&&\\ \hline
   
 &x= 0.30 &&&\\
 $E_n$  &  -1.448782 & 0.906504&  5.254129&  8.288149\\
 $(n|P|n)$ &     0.868986 & 0.460296 & 0.566014&  0.104704\\
 $E_{LS}$ & -1.620734 & 0.339561&& \\
 $E_{KK}$  &  -1.125579 & 5.515468&&\\
 $E_{EKKO}$ & -1.448782 & 0.906504&&\\ \hline \hline 
   
\end{tabular}
\end{center}
\end{table}

\begin{figure}
\scalebox{0.42}{
\includegraphics[angle=-90]{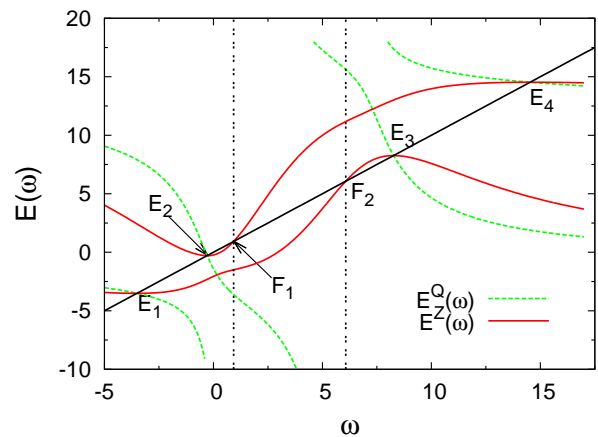}
}
\caption{Graphical solutions for the $\hat Q$- and $\hat Z$-box 
self-consistent equations. See text for other explanations.}
\end{figure}

\subsection{IIIb. The $sd$ and $sdpf$ shell model effective 
interactions}

In this subsection, we shall  calculate the effective interactions
for both the degenerate  $sd$ one-shell and the non-degenerate $sdpf$
two-shell cases. Before presenting our results, let us first
describe some details of our calculations. The LS, KK and EKKO methods as
 described in section II will be employed. We first 
compute the low-momentum nucleon-nucleon interaction
$V^{2N}_{low-k}$ 
\cite{bogner01,bogner02,bogner03,coraggio09}
starting from the chiral N$^3$LO two-body potential \cite{idaho}
at a decimation scale 
of $\Lambda =$ 2.1 fm$^{-1}$. At this cutoff scale, the low-momentum 
interactions derived from
different NN potentials \cite{idaho,cdbonn,argonne,nijmegen}
are remarkably close to each other, leading to a nearly unique
low-momentum interacton \cite{bogner03}. 
The effect of the leading-order chiral three-nucleon force on 
shell model effective interactions has been studied, and our 
results will be reported in a separate publication \cite{dongv3n}.
The above $V^{2N}_{low-k}$ 
interaction is then used in calculating the $\hat Q$-box diagrams
as shown in Fig.\ 1. In calculating these diagrams, the hole orbits
are summed over the $0s0p$ shells and particle orbits over the $0d1s1p0f$
shells. The active spaces ($P$-space) used for the one-shell and 
two-shell calculations are respectively
 the  three orbits in the $sd$ shell and the seven orbits in the $sdpf$
shells. The experimental s.p.\ energies of (0.0, 5.08, 0.87) MeV
have been used, respectively, for the 
($0d_{5/2},0d_{3/2},1s_{1/2}$) orbits  \cite{nucldata}.
We have employed the shell-model s.p.\ wave functions and energies with
the harmonic oscillator constant of $\hbar \omega$=14 MeV.  

\begin{figure}
\scalebox{0.42}{
\includegraphics[angle=-90]{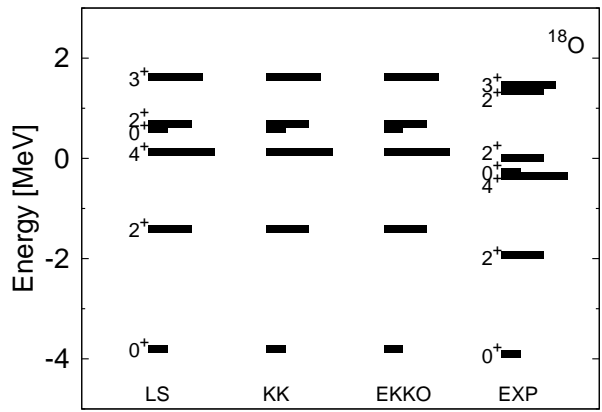}
}
\caption{Energy spectrum of $^{18}$O obtained from the $sd$-shell 
calculations using the LS, KK and 
EKKO methods. See text for more explanations. }
\end{figure}

\begin{figure}
\scalebox{0.42}{
\includegraphics[angle=-90]{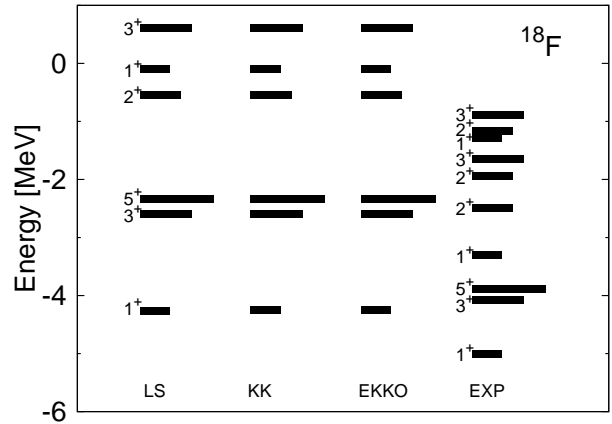}
}
\caption{Same as Fig.\ 3 except for $^{18}$F. }
\end{figure}
In the following, let us first report our results for the degenerate $sd$
one-shell calculations. Here by degenerate we mean that the 
unperturbed s.p.\ energy
levels for the ($0d_{5/2},~0d_{3/2},1s_{1/2}$) orbits are degenerate.
Our purpose here is mainly to compare
the results given by the KK and EKKO method with those given by the commonly
used degenerate LS method \cite{jensen95,coraggio09}. 
Our results are presented in  Figs.\ 3 and 4.
As discussed in section II and illustrated in Table I, 
the  EKKO and KK methods may converge to different states. In the present
calculations they actually converge to the same states, 
as seen in Figs.\ 3 and 4.
 Our LS calculations are carried out using a low-order 
approximation, namely we take $V_{eff}=R_5$ (see eq.\ (7)).
As shown in the figures, the LS, KK and EKKO results
for both $^{18}$O and $^{18}$F 
are in fact nearly identical to each other. 
A comparison of our results with experiments \cite{nucldata} is also
presented in the figures. The agreement between our calculated energy
levels with experiments is moderately  satisfactory for $^{18}$O, but for
$^{18}$F the calculated lowest $(1^+,3^+,5^+)$ states, though of
correct ordering, are all significantly higher than the experimental
values. 
 As discussed in sections II and IIIa, 
the LS method is known to converge to the
states of the lowest energies, while the KK method to the states 
of maximum $P$-space overlaps. Thus the good agreement shown  in
Figs.\ 3 and 4 is an important indication that the states reproduced by the 
model-space LS, EKKO and KK effective
interaction are likely those of the lowest energies as well as maximum
model-space overlaps. 
In Fig.\ 5, we compare the entire
$sd$-shell matrix elements given by the LS and EKKO  methods;
 it is remarkable that every individual LS matrix element  is  
practically equal to the corresponding EKKO one.
(The matrix elements given by EKKO and KK are also nearly identical.) 
Recall that our LS matrix
elements were obtained with a low-order $R_5$ iteration, which indicates 
the rapid convergence of the iteration scheme in the case of $sd$-shell 
effective interactions.

 As discussed in sections II and IIIa, 
the EKKO or KK methods are convenient
for deriving the effective interactions for non-degenerate model spaces. 
In the following let us first apply these methods to a relatively simple
case, namely the non-degenerate
$sd$ one-shell effective interactions. The shell-model s.p.\ energies
for the $sd$ shell are degenerate, but the experimental ones are not.
It may be of interest to employ the experimental s.p.\ energies
as the unperturbed s.p.\ energies for the $sd$ shell \cite{coraggio09}.
Thus here we employ the same unperturbed s.p.\ spectrum
as the previous degenerate $sd$ case except that
the $0d_{3/2}$ and $1s_{1/2}$ orbits are shifted upward by, respectively,
5.08 and 0.87 MeV relative to the $0d_{5/2}$, mimicking the experimental
s.p. energies.  We have found that 
the non-degenerate and degenerate 
$sd$ EKKO effective interactions  do not differ significantly
from each other. 
To illustrate,
we compare in Fig.\ 6 the spectra of $^{19}$F 
for both the degenerate and non-degenerate $sd$-shell calculations;
they agree with each other rather well. Slightly better
 agreements between the
spectra of $^{18}O$, $^{18}F$ and $^{19}O$ calculated with 
these two interactions are also observed.
In short, the one-shell
$sd$ effective interactions given by the degenerate and non-degenerate
choices for the unperturbed $sd$ s.p.\ energies are nearly the same,
and it is adequate in this case to just use the former choice.

\begin{figure}
\scalebox{0.42}{
\includegraphics[angle=-90]{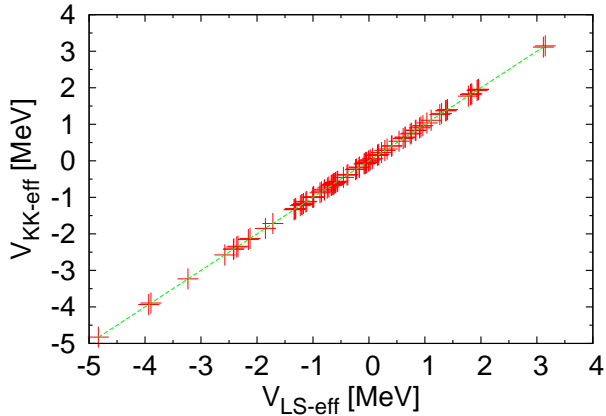}
}
\caption{Comparison of the $sd$-shell matrix elements given
by the LS (LS-eff) and EKKO (KK-eff) methods. 
The diagonal line indicates the equivalence
of the LS and EKKO matrix elements. See text for more explanations.}
\end{figure}

\begin{figure}
\scalebox{0.42}{
\includegraphics[angle=-90]{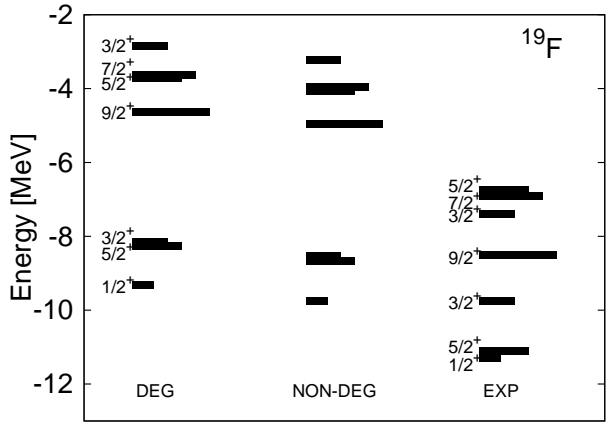}
}
\caption{Comparison of the low-lying states of $^{19}$F calculated
with the degenerate (DEG) and non-degenerate (NON-DEG) $sd$ one-shell 
EKKO effective interactions.}
\end{figure}

\begin{figure}
\scalebox{0.42}{
\includegraphics[angle=-90]{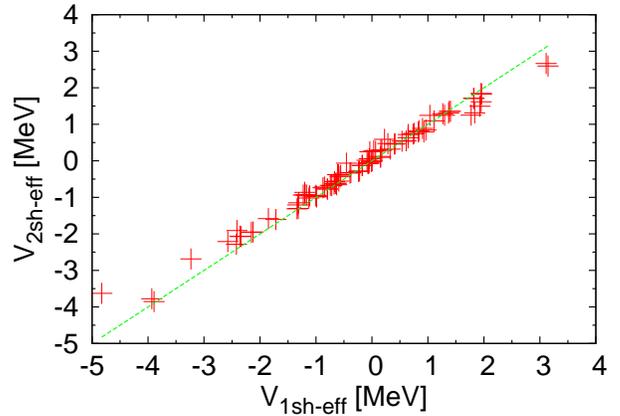}
}
\caption{Comparison of the matrix elements of the $sd$ one-shell (1-eff)
and $sdpf$ two-shell (2-eff) EKKO effective interactions. 
Only the matrix elements
within the $sd$ shell are shown. See text for more
explanations.}
\end{figure}

\begin{figure}
\scalebox{0.42}{
\includegraphics[angle=-90]{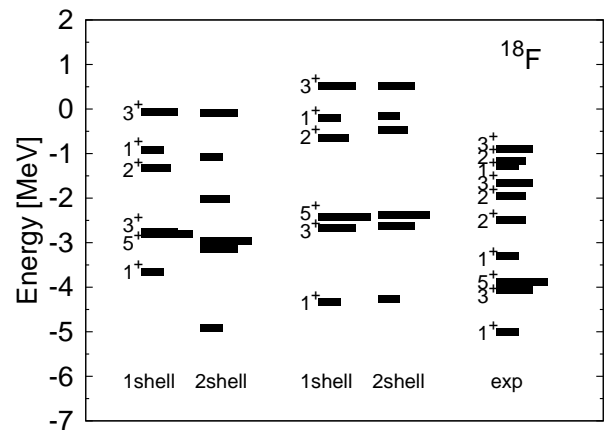}
}
\caption{Low-lying states of $^{18}$F calculated with the $sd$ one-shell
(1shell) and $sdpf$ two-shell (2shell) EKKO interactions. 
Two choices for the vertex function
$\hat Q$-box are employed.  For the left-most two columns  only 
the first-order $\hat Q$-box diagrams are included, while for the 
middle two columns
 we include both the first- and second-order diagrams.}
\end{figure}

We now report our results for 
the non-degenerate $sdpf$ two-shell effective interactions.  
So far our calculations   
have all been carried out
using the $sd$ one-shell model space. For certain nuclei such as those
with a large neutron excess, a larger model space such 
as the $sdpf$ one may be needed.
It will be convenient to describe our $sdpf$ calculations 
by way of an example, namely $^{18}$O.
Consider the $(T=1,J=0)$ states of this nucleus. In the $sd$ one-shell
case, the model space is spanned by three basis states
$|j^2,T=1,J=0\rangle$, where $j=d_{5/2},d_{3/2}$, or $s_{1/2}$. For the $sdpf$ case,
the model space is enlarged, having four additional single-particle states
labeled by $j=(p_{3/2},p_{1/2},f_{7/2},f_{5/2}$). The whole 
model space is now non-degenerate as the $pf$ orbits 
are one shell above the $sd$ ones. 
Similar to the $sd$ one-shell case, we have calculated 
the $sdpf$ effective interactions
using both the EKKO and KK methods. 
For the $sd$ shell-model calculations, we have employed the experimental
s.p.\ energies for the three $sd$-shell orbits as mentioned earlier. 

 For the $sdpf$ shell-model calculations, we need in addition 
the experimental
s.p.\ energies for the four $pf$ orbits. Their values are, however,
not well known. In the present calculation we have placed them all at 
a separation of one
$\hbar \omega$ (14 MeV) above the $d_{5/2}$ level. 
 (As to be reported later (Fig. 11),
we shall also use a smaller value for the above separation.) 
In our calculations we have considered two choices for the unperturbed
s.p.\ energies of the $sd$ model space, a degenerate one 
and a non-degenerate one
(in which the  $0d_{3/2}$  and $1s_{1/2}$ orbits are shifted higher
in energy compared to the $0d_{5/2}$ orbit as described earlier). 
We have found that the results are rather similar, and in the following
discussion we report only the calculations for
the degenerate $sd$-shell choice. 

In Fig.\ 7 we compare the matrix elements of the $sdpf$ two-shell
EKKO interaction
with those of the $sd$ one-shell case. Only the matrix elements
within the $sd$ shell are shown. We find that the magnitudes of the
two-shell matrix elements are generally weaker than the one-shell
matrix elements, some differing by as much as 1 MeV.
Despite these differences, the resulting spectra
for $^{18}$O and $^{18}$F given by the one-shell and two-shell
calculations are nearly equivalent to each other, as illustrated
by the two middle columns of Fig.\ 8.
In our present and subsequent calculations we employ 
a low-order $\hat Q$-box
consisting of the  1st- and 2nd-order diagrams of Fig.\ 1. It 
is instructive, however, to compare the one- and two-shell calculations
when only the leading-order approximation to the $\hat Q$-box is
retained.
In the two leftmost columns of Fig.\ 8 we show the spectra of 
$^{18}$F when only the 1st-order $\hat Q$-box diagrams are included. 
We note that in this case (in contrast to the second-order $\hat Q$-box 
calculation) the resulting $sd$ 
`1 Shell' and $sdpf$ `2 Shell' spectra are largely different.

 This important observation can be explained as follows. In general, the 
different $sd$ and $sdpf$ model spaces will result in different
effective interactions and associated effective Hamiltonians, 
which we denote by $H_{eff}^1$ and $H_{eff}^2$.
But since the $sd$ model space is a subspace of $sdpf$ model space, 
$H_{eff}^1$ and $H_{eff}^2$ should in principle have common 
eigenvalues. We would like to check that this requirement
is satisfied by our present calculations.
As indicated by the two middle columns of Fig.\ 8, we see that 
indeed this is the case.
In fact among the many states given by the
$sdpf$ calculations, it is the ones with the maximum $sd$-space overlaps
which agree with the results given by the $sd$-shell calculation, which is
a physically desirable result. 
Because of the difference in the model spaces, the renormalization effects
for the $sd$ and $sdpf$ effective interactions are different.
To bridge these differences, we need to include
at least the 2nd-order $\hat Q$-box diagrams
(note that the allowed intermediate
states of the 2nd-order $\hat Q$-box diagrams are model-space dependent).
These diagrams are not included in the above 1st-order calculations,
and consequently the $sd$ and $sdpf$ results are different
as shown by the two leftmost columns of Fig.\ 8, which
is a strong evidence for the importance of the 
model-space-dependent renormalization effects. The construction of
model-space effective interactions is in many ways similar to the
construction of an effective theory from a renormalization group 
evolution. In such cases, one would expect that despite the different 
effective Hamiltonians, the same low-energy physical observables would
be reproduced. The shell model effective interactions in the present 
work have not been computed exactly (that is, including high-order
diagrams in the $\hat Q$-box), yet it is interesting that we
nevertheless find excellent agreement between the one- and
two-shell calculations including $\hat Q$-box diagrams only up to 
second order.

\begin{figure}
\scalebox{0.42}{
\includegraphics[angle=-90]{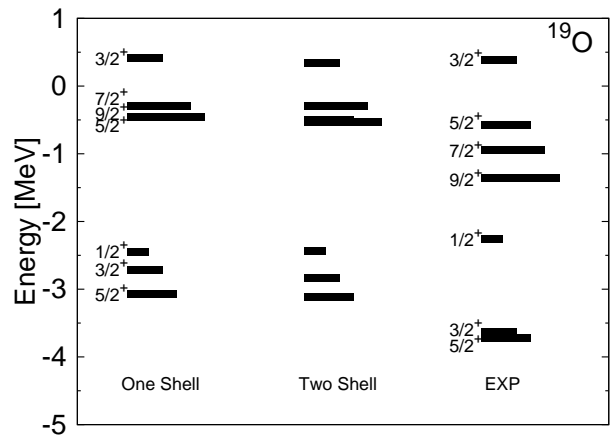}
}
\caption{Low-lying states of $^{19}$O calculated with the $sd$ one-shell
(1 Shell) and $sdpf$ two-shell (2 Shell) EKKO interactions. The experimental
results are from \cite{nucldata}.}
\end{figure}

\begin{figure}
\scalebox{0.42}{
\includegraphics[angle=-90]{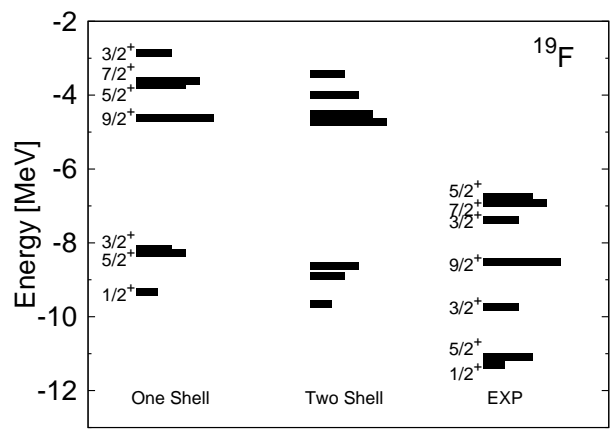}
}
\caption{Same as Fig.\ 10 except for $^{19}$F.}
\end{figure}

\begin{figure}
\scalebox{0.42}{
\includegraphics[angle=-90]{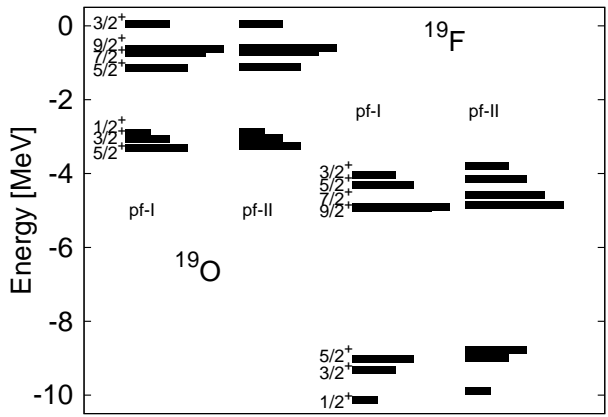}
}
\caption{Low-lying states of $^{19}$O and $^{19}F$ 
calculated using the  $sdpf$ two-shell 
 EKKO interactions,  with the $pf$ experimental s.p. energies 
placed at 10  ('pf-I') and 14 MeV ('pf-II') above the $0d_{5/2}$
orbit.}
\end{figure}

\begin{figure}
\scalebox{0.30}{
\includegraphics[angle=0]{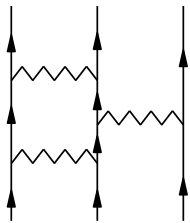}
}
\caption{A shell-model three-nucleon force diagram. Its external lines
all belong to the $sd$ shell, while its intermediate states
between any two vertices must have at least one line 
belonging to the $pf$ shell.}
\end{figure}

 To further study the $sd$ one-shell and $sdpf$ two-shell effective
interactions, we have applied them to shell model calculations 
of $^{19}$O and $^{19}$F. The 2nd-order $\hat Q$-box of Fig.\ 1 is employed.
Our results are displayed respectively in Figs.\ 9 and 10.
 It is of interest that the one-shell and two-shell results
for $^{19}$O are nearly identical, and for $^{19}$F they are also
remarkably close to each other. Although the orderings of our
calculated spectra are in fair agreements with experiments, there
are significant differences between them.  
 As discussed earlier, in our $sdpf$ two-shell calcualtions
the experimental s.p. energies for the four $pf$ orbits are needed, but their
values are not well known.  So far we have chosen to place them
at a separation of 14 MeV above the $0d_{5/2}$ orbit. We have repeated
our calculations using instead a smaller separation of 10 MeV, to
investigate if the use a smaller separation may improve
the agreements. As illustrated in Fig. 11, the low-lying states of $^{19}O$
are hardly changed, those given by the separations of 14 and 10 MeV being
 nearly identical. The differences between the two sets of states for $^{19}F$ 
are also generally small.
That these low-lying states are insensitive to the above separations 
 indicates that our $sdpf$ effective interactions
(obtained with the inclusion of the 1st- and 2nd-order $\hat Q$-box 
diagrams of Fig. 1) have a rather weak coupling between the $sd$ and
$fp$ shells.

Recall that we have employed the folded-diagram
expansion of eq.\ (1) to calculate the effective interaction $V_{eff}$.
For nuclei with three valence nucleons, this expansion has both 2-body
and 3-body diagrams. (By 3-body diagrams we mean those valence-linked 
diagrams with three incoming and three outgoing valence lines \cite{ko90}.)
As an example, the shell-model three-nucleon force
diagram of Fig.\ 12 should be included in the $sd$ one-shell effective
interaction for $^{19}$O and $^{19}$F. But this diagram is not
included in our present $sd$ one-shell calculation as we employ only
the two-body $\hat Q$-box of Fig.\ 1.  This 
diagram is, however,
included in our $sdpf$ shell model calculations for $^{19}$O
and $^{19}$F.  Thus the good agreement
between the one-shell and two-shell results shown in Fig.\ 9 is 
an indication that this type of shell-model
three-nucleon force is likely to have little importance for $^{19}$O
spectra,
while it is moderately important for $^{19}$F as suggested by 
the small difference shown in Fig.\ 10. 
Further studies of this
type of three-nucleon force will be useful and of interest, and we plan
to do so in a future study.  

\section{IV. Summary and conclusion}

We have applied the iteration method of Krenciglowa and Kuo (KK) 
and that recently
developed by Okamoto {\it et al}.\ (EKKO) to the microscopic
derivation of the $sd$ and $sdpf$ shell-model effective interactions 
using the low-momentum nucleon-nucleon interaction derived from
the chiral $N^3LO$ two-body potential.
We first considered a solvable model and found that
both methods are suitable and efficient for deriving the effective interactions
for non-degenerate model spaces, where the Lee-Suzuki iteration method
is considerably less convenient. Even in the situation where 
the $P$- and $Q$-space unperturbed Hamiltonians have spectrum overlaps 
did the KK and EKKO methods perform remarkably well.
The EKKO method 
has the special advantage that its vertex function $\hat Z$-box
is, by construction, a continuous function of the energy, while the $\hat Q$-box
function used in the LS and KK methods may have singularities. This 
feature was found to be particularly useful for the convergence of the EKKO 
iteration method.
 
Using the $V_{low-k}$ low-momentum potentialss obtained from
above two-body interaction, we first 
calculated the degenerate $sd$ one-shell effective
interactions using the LS, KK and EKKO methods. The results given by KK and 
EKKO were found to be identical. It is noteworthy 
that the LS results, calculated
with a low-order (5th order) iteration, were also in very good agreement with
both the KK and EKKO results, supporting the accuracy of the low-order
LS method for calculating the degenerate shell-model effective
interactions. 

We have calculated the non-degenerate $sdpf$ two-shell 
effective interactions using both the EKKO and KK methods. 
Both methods gave identical results and were found to be suitable
for such non-degenerate calculations, with the former
being more efficient (faster converging).
We have applied these interactions
to compute the low-lying energy spectra for several nuclei with two 
and three valence nucleons above the $^{16}$O core. 
Since the $sd$ model space is a subspace of the $sdpf$ model space, 
we expect the effective 
Hamiltonians for these two spaces to have common
eigenvalues. Indeed this was largely confirmed in our calculations of
$^{18}$O, $^{18}$F, $^{19}$O and $^{19}$F spectra, where it was found that
the states in the $sdpf$ calculations with the maximum $sd$-space overlap
agreed with the results given by the $sd$ calculations. 
The above agreement was found to be excellent for $^{19}$O, though
not as good for $^{19}$F, which indicates that the shell-model three-nucleon
force is more important in $^{19}$F (where the proton-neutron interaction 
is involved) than in $^{19}$O.
Further study of this three-nucleon force should be useful
and of much interest.

The calculated ground state 
energies for the above four nuclei are all higher (less bound)
than the corresponding experimental values. We are studying if the
inclusion of the chiral three-nucleon force may give additional
binding energy \cite{dongv3n}. In the present work
we have employed the $\hat Q$-box irreducible vertex function
consisting of the first- and second-order diagrams only. The inclusion
of the Kirson-Babu-Brown (KBB) all-order core polarization diagrams in the 
$\hat Q$-box may provide additional binding energy \cite{jwholt05}.
We plan to carry out futher calculations with the inclusion
of such all-order KBB diagrams.

\begin{acknowledgments}
We are very grateful to L.\ Coraggio, A.\ Covello, 
 A.\ Gargano,  Jason Holt,  N.\ Itaco, M.\ Machleidt, 
R.\ Okamoto and K.\ Suzuki  for many 
helpful discussions. 
Partial supports from the US Department of Energy under contracts
DE-FG02-88ER40388 and the DFG (Deutsche Forschungsgemeinschaft)
 cluster of excellence: Origin and Structure of the Universe are
 gratefully acknowledged.
\end{acknowledgments}

\end{document}